\documentclass[11pt]{article}

\usepackage{axodraw}
\usepackage{epsfig,graphicx,psfrag} % Include figure files
\usepackage{amssymb}
\usepackage{cite}

\newcommand{\be}{\begin{equation}}
\newcommand{\ee}{\end{equation}}
\newcommand{\bear}{\begin{eqnarray}}
\newcommand{\eear}{\end{eqnarray}} 

\newcommand{\one}{{\prime} }

\begin{document}

\twocolumn[ 

{\scriptsize FERMILAB-PUB-07-459-T} \\  

\vspace{-1.1cm}
\title{\bf \Large Massive color-octet bosons and pairs of resonances at hadron colliders}
\author{\large Bogdan A. Dobrescu, Kyoungchul Kong, Rakhi Mahbubani \\ [4mm]  
\it Theoretical Physics Department, Fermilab, Batavia, IL 60510, USA }
\date{\small July 1, 2008; revised October 16, 2008} 
\maketitle

\vspace{-11mm}

\begin{quote}
We analyze collider signatures of massive color-octet bosons whose couplings 
to quarks are suppressed. Gauge invariance forces the octets to couple at 
tree level only in pairs to gluons, with a strength set by the QCD gauge 
coupling. For a spin-1 octet, the cross section for pair production 
at hadron colliders is larger than that for 
a quark of equal mass. The octet decays into two jets, leading to a 4-jet 
signature with two pairs of jets forming resonances of the same mass.
For a spin-0 octet the cross section is smaller, and the dominant decay is into $b\bar{b}$, 
or $t\bar{t}$ if kinematically allowed. 
We estimate that discovery of spin-1 octets is possible for masses 
up to 330 GeV at the Tevatron, and 1 TeV at the LHC with 1 fb$^{-1}$,
while the reach is somewhat lower for spin-0 octets. \\
\end{quote}

%\pacs{12.60.-i, 13.85.-t, 14.80.-j}
]

%\bigskip\bigskip
\vspace{-2mm}
%%%%%%%%%%%%%%%%%%%%%%%%%%%%%%%%%%%%%%%%%%
%%%%%%%%%%%%%%%%%%%%%%%%%%%%%%%%%%%%%%%%%%%%%%%%%%%%%%%%%%%%%%%%%%%%%%%%%
%{\it Introduction.}---

\section{ \large  Introduction}
%\section{Introduction}

The first discovery of physics
beyond the standard model could consist of signals from an effective
theory that includes a single new particle. 
The number of interesting theories of this type is limited because particles 
are identified by a few quantum numbers (especially spin and gauge charges)
which take only a small number of discrete values.
Well-known examples include $Z^\prime$ bosons, vectorlike quarks 
or singlet scalars.

In this letter we study a type of particle present in a variety
of theories, whose collider signatures have been less intensely investigated:
massive color-octet vector bosons. 
These arise from an $SU(3)_1\times SU(3)_2$ gauge group
spontaneously broken down to its diagonal subgroup $SU(3)_c$,
which is identified with the QCD gauge symmetry. Such a pattern of gauge
symmetry breaking, earlier invoked for phenomenological or aesthetic
reasons \cite{Preskill:1980mz}, has been used 
in topcolor models of Higgs compositeness \cite{Hill:1991at,Hill:2002ap},
and was one of the motivations for studying models with extra dimensions 
\cite{Dobrescu:1998dg}.
Color-octet bosons may also be composite particles, such as one of the 
$\rho$ techni-mesons \cite{Hill:2002ap}.
Given that such particles may be associated with many 
extensions of the standard model, we explore their properties 
independently of the underlying theory that might justify their existence.
In particular, we show that vector octets can give rise to striking multi-jet signatures
at hadron colliders and that these could be experimentally observed even within 
existing data sets once a dedicated analysis is performed.

Distinguished by its coupling to quarks, the spin-1 color octet has been called
a coloron (vector coupling \cite{Hill:1993hs,Chivukula:1996yr,Choudhury:2007ux}), 
an axigluon (axial vector coupling \cite{Bagger:1987fz,Choudhury:2007ux}), or
a topgluon (preferential coupling to top quarks \cite{Hill:2002ap,Choudhury:2007ux},
which is also the case for Kaluza-Klein gluons from a warped 
extra dimension \cite{Lillie:2007yh}). 
This illustrates the model dependence of couplings between quarks and 
color-octet vector bosons, which may even vanish at tree-level, as in the case of 
level-1 gluons in models with universal extra dimensions.
We will present some simple 4-dimensional extensions of the standard model 
in which such couplings are suppressed, and 
as a result single production of the color-octet boson is small. 

We focus on pair production of spin-1 octets at hadron colliders,
which has a large and mostly model-independent rate  (see also \cite{Dicus:1994sw,Zerwekh:2006te}).
The main decay mode is into two quarks, 
so that the signature is a pair of dijet resonances, for which the backgrounds 
are highly reducible. 
Spin-0 octets have similar properties \cite{Eichten:1984eu,Chivukula:1991zk,Hill:2002ap}, 
with the distinction that 
their couplings to quarks are generically proportional to the quark mass. 
We study the prospects for observing these particles at the Tevatron and the LHC.

%\smallskip
%%%%%%%%%%%%%%%%%%%%%%%%%%%%%%%%%%%%%%%%%%%%%%%%%%%%%%%%%%%%%%%%%%%%%%%%%
%{\it Interactions of the vector octet.}---

\section{\large Interactions of the vector octet}

Consider an effective field theory that includes a vector
color-octet field $G_\mu^\one$ of mass $M_G$, where $\mu$ is a Lorentz index.
Any massive spin-1 particle may be identified with the gauge boson 
of a spontaneously broken gauge symmetry, provided
higher-dimensional operators are included. 
We assume that this effective theory is valid over a range of scales above $M_G$, 
so that the higher-dimensional operators are suppressed and may be neglected.
The gauge symmetry breaking pattern that 
gives only a massive spin-1 octet and the massless gluon
is $SU(3)_1 \times SU(3)_2 \to SU(3)_c$. 
Any other gauge group that gives rise to $G_\mu^\one$ should embed this 
minimal gauge group.
%heavy spin-1 particle 
The interactions of the massive octet with gluons may be derived 
by rotating the $SU(3)_1$ and $SU(3)_2$ gauge kinetic terms 
to the mass eigenstate basis. %Only a
A pair of $G_\mu^\one$ bosons 
couples at tree level to one or two gluons: 
\begin{eqnarray}
&& \hspace*{-0.8cm} \frac{g_s^2}{2}f^{abc}f^{ade} \, G^{\one \mu b} 
\left[ G^{\nu d}\left(G_\nu^{\one c} G_\mu^{e} + G_\mu^{\one e} G_\nu^{c}\right) 
+ G^{\one e}_\nu G^{\nu c} G_\mu^{d}\right] \nonumber \\
&& \hspace*{-0.7cm} + \, g_sf^{abc}  G_\mu^{\one a} \left[\left(\partial^\mu G^{\one \nu b}
- \partial^\nu G^{\one \mu b}\right) G_\nu^{c}
- G_\nu^{\one b}\, \partial^\mu G^{\nu c}\right] ~.  \nonumber \\ 
\label{self}
\end{eqnarray}
Here $g_s$ is the QCD gauge coupling, 
$f^{abc}$ are the $SU(3)_c$ structure constants, and $G_\mu$ is the gluon field.
The above interactions have an accidental $Z_2$ symmetry because $G_\mu^\one$ appears 
only in pairs. 
%The $G_\mu^\one$ boson may also couple to gluons through operators of dimension-6 or higher,
%but we assume that such operators are suppressed by a scale sufficiently higher than $M_G$.

The most general Lorentz-invariant dimension-4 interactions of $G_\mu^\one$ with quarks are
of the type $G_\mu^{\one a} \overline{q}\gamma^\mu T^a q^\prime$,
where $q$ and $q^\prime$ are quarks carrying the same electroweak charges, and $T^a$ are the 
generators of the fundamental representation of $SU(3)_c$.
The coefficients of these operators form three different $3\times3$ Hermitian matrices.
To avoid a lengthy discussion of
flavor-changing processes, we assume that these three matrices have 
diagonal elements approximately equal (up to a sign) to a parameter $h_q > 0$,
and negligible off-diagonal elements.
There is an upper limit on $h_q$ set by dijet searches.
For a $G_\mu^\one$ mass $M_{G}$ between 150 and 200 GeV the
limit is at most $h_q < g_s/4$ \cite{Albajar:1988rs}, while for some values of $M_{G}$
above 200 GeV the limit is more stringent, around $h_q < g_s/7$ \cite{Abe:1997hm}.

Such suppressed couplings of $G_\mu^\one$ %spin-1 color octets 
to quarks can arise in simple renormalizable models.
For example, let us consider an $SU(3)_1\times SU(3)_2$ gauge theory 
where the breaking down to $SU(3)_c$ is due to 
the vacuum expectation value of a complex scalar field $\Phi$ (or
of a fermion-antifermion pair, induced by some technicolor-like interaction),
which transforms as a bifundamental under the product gauge group.
After diagonalizing the gauge boson mass matrix, the 
massless gluon has a gauge coupling $ g_s = h_1h_2/\sqrt{h_1^2 + h_2^2}$,
where $h_1$ and $h_2$ are the $SU(3)_1\times SU(3)_2$ gauge couplings.
This is to be identified with the QCD coupling at the scale $M_G$:
$g_s \approx 1.1$ for $M_G$ of a few hundred GeV.
Imposing perturbativity of both $SU(3)$ interactions  
at the symmetry breaking scale gives $g_s < h_1,h_2 \lesssim \sqrt{4\pi}$. 
In the gauge eigenstate basis, the quarks that are triplets under $SU(3)_1$ couple to 
$G_\mu^\one$ with a strength $h_q = g_s h_1/h_2$,
with  $h_1$ and $h_2$  interchanged for triplets under $SU(3)_2$.
 A simple choice is that all observed quarks transform as triplets under 
$SU(3)_1$ \cite{Chivukula:1996yr}. By itself, this would lead to a large coupling 
of $G_\mu^\one$ to quarks, $h_q \gtrsim g_s^2/\sqrt{4\pi} \approx 0.3$, which would
imply that most $G_\mu^\one$ masses between 250 and 750 GeV are ruled out by the CDF 
search \cite{Abe:1997hm}.
However, in the presence of new heavy quarks which mix with the observed ones,
the couplings of  $G_\mu^\one$ may change dramatically. Consider a vectorlike 
quark $Q$ whose left- and right-handed components transform as a $3$ under 
$SU(3)_2$, and like standard model left-handed quarks ($q_L$) under $SU(2)_W \times U(1)_Y$. 
In addition to a mass term for $Q$,
the Lagrangian includes Yukawa couplings of the vectorlike quark to $q_L$ and $\Phi$.
The off-diagonal mass term induced by $\langle \Phi \rangle$ requires a rotation of $q_L$ and $Q_L$
by an angle $\theta$, such that  the coupling of $G_\mu^\one$ to $\overline{q}_L\gamma^\mu T^a q_L$ 
becomes
\bear
h_q = g_s \left(\frac{h_1}{h_2} \cos^2\!\theta - \frac{h_2}{h_1} \sin^2\!\theta\right) ~,
\eear
and an ``off-diagonal'' interaction $G_\mu^{\one a}\overline{Q}_L\gamma^\mu T^a q_L$ is induced.
For $\tan\theta = h_1/h_2$, we find $h_q=0$, while the coefficient of the 
off-diagonal interaction becomes $g_s$.
By including a vectorlike quark for each standard model quark,
one may in principle arrange that all tree-level couplings of  $G_\mu^\one$ to standard model 
currents vanish. Such cancellations require fine-tuning, and are unlikely to be realized precisely in nature. 
Nevertheless, cancellations at the  15\% level are sufficient to free 
$G_\mu^\one$ from the existing limits on dijet resonances.
Note that the mixings between standard model quarks and vectorlike ones may be approximately
flavor independent, so that the induced flavor-changing neutral 
currents are not necessarily large.

A more sophisticated model includes an additional $SU(3)$ gauge group, and invariance under 
a $Z_2$ symmetry that interchanges two of the groups. There are in this case two heavy spin-1 
octets. By virtue of the $Z_2$ symmetry, the couplings of the lighter octet to standard model 
quarks vanish exactly. The heavier octet has sizable couplings to quarks 
in the gauge eigenstate basis,
but as in the previous model, in the presence of some vectorlike quarks the couplings to 
mass eigenstates may be partially canceled. The field content of this model 
resembles the colored Kaluza-Klein modes of the first two levels in theories 
with universal extra dimensions.

%\smallskip
%%%%%%%%%%%%%%%%%%%%%%%%%%%%%%%%%%%%%%%%%%%%%%%%%%%%%%%%%%%%%%%%%%%%%%%%%
%{\it $G_\mu^\prime$ production at hadron colliders.}---

\section{\large  $G_\mu^\prime$ production at hadron colliders}

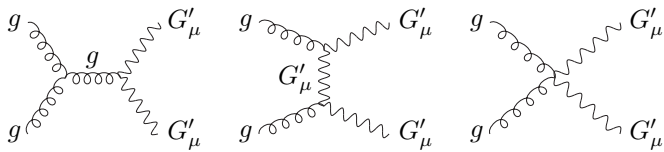
\begin{figure}[b!]
\vspace*{-0.1cm}
\begin{center} 
{
\unitlength=0.7 pt
\SetScale{0.7}
\SetWidth{0.5}      % line    size control
\normalsize    %  letter  size control
{} \allowbreak
%  diagram # 1
\begin{picture}(100,100)(8,0)
\Gluon(15,80)(35,50){3}{4}
\Gluon(15,20)(35,50){3}{4}
\Gluon(35,50)(65,50){3}{4}
\Photon(65,50)(85,80){3}{5}
\Photon(65,50)(85,20){3}{5}
%\Text(50,10)[c]{(a)}
\Text(50,60)[c]{\small $g$}
\Text( 8,80)[c]{\small $g$}
\Text( 8,20)[c]{\small $g$}
\Text(100,80)[c]{\small $G_\mu^\prime$}
\Text(100,20)[c]{\small $G_\mu^\prime$}
\end{picture}
\quad
%
%  diagram # 2
\begin{picture}(100,100)(4,0)
\Gluon(15,80)(50,65){3}{4}
\Gluon(15,20)(50,35){3}{4}
\Photon(50,65)(50,35){3}{5}
\Photon(50,65)(85,80){3}{5}
\Photon(50,35)(85,20){3}{5}
%\Text(50,10)[c]{(b)}
\Text(35,50)[c]{\small $G_\mu^\prime$}
\Text( 8,80)[c]{\small $g$}
\Text( 8,20)[c]{\small $g$}
\Text(100,80)[c]{\small $G_\mu^\prime$}
\Text(100,20)[c]{\small $G_\mu^\prime$}
\end{picture}\
\quad
%
%  diagram # 3
\begin{picture}(100,100)(0,0)
\Gluon(15,80)(50,50){3}{4}
\Gluon(15,20)(50,50){3}{4}
\Photon(50,50)(85,80){3}{5}
\Photon(50,50)(85,20){-3}{5}
%\Text(50,10)[c]{(c)}
\Text( 8,80)[c]{\small $g$}
\Text( 8,20)[c]{\small $g$}
\Text(100,80)[c]{\small $G_\mu^\prime$}
\Text(100,20)[c]{\small $G_\mu^\prime$}
\end{picture}
}
\end{center}
\vspace*{-0.8cm}
\caption{$G_\mu^\one G_\mu^\one$ production from $gg$
initial state ($u$-channel $G_\mu^\one$ exchange is not shown). 
Curly lines represent gluons, 
while wavy lines represent massive vector octets.} 
\label{fig:gg}
\end{figure}

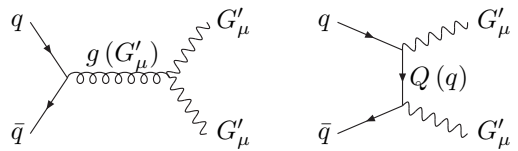
\begin{figure}[b!]
\vspace*{-0.1cm}
\begin{center} 
{
\unitlength=0.7 pt
\SetScale{0.7}
\SetWidth{0.6}      % line    size control
\normalsize    %  letter  size control
{} \allowbreak
%  diagram # 4
\begin{picture}(100,100)(0,0)
\ArrowLine(-10,80)(10,50)
\ArrowLine(10,50)(-10,20)
\Gluon(10,50)(65,50){3}{7}
\Photon(65,50)(85,80){3}{5}
\Photon(65,50)(85,20){3}{5}
%\Text(50,10)[c]{(d)}
\Text(40,62)[c]{\small $g\, (G_\mu^\prime)$}
\Text( -17,80)[c]{\small $q$}
\Text( -17,20)[c]{\small $\bar{q}$}
\Text(100,80)[c]{\small $G_\mu^\prime$}
\Text(100,20)[c]{\small $G_\mu^\prime$}
\end{picture}\
\quad
%
%  diagram # 5
\begin{picture}(100,100)(-20,0)
\ArrowLine(15,80)(50,65)
\ArrowLine(50,35)(15,20)
\ArrowLine(50,65)(50,35)
\Photon(50,65)(85,80){3}{5}
\Photon(50,35)(85,20){3}{5}
%\Text(50,10)[c]{(e)}
\Text(70,50)[c]{\small $Q\, (q)$}
\Text( 8,80)[c]{\small $q$}
\Text( 8,20)[c]{\small $\bar{q}$}
\Text(100,80)[c]{\small $G_\mu^\prime$}
\Text(100,20)[c]{\small $G_\mu^\prime$}
\end{picture}\
\quad
}
\end{center}
\vspace*{-0.9cm}
\caption{$G_\mu^\one G_\mu^\one$ production  from $q\bar{q}$ initial state ($u$-channel
diagram is not shown).
If couplings of standard model quarks ($q$) to $G_\mu^\one$ are
suppressed due to mixing with vectorlike quarks ($Q$), then 
$G_\mu^\one$ or $q$ exchange  contributions may be negligible.
}
\label{fig:qqbar}
\end{figure}

The Feynman rules 
 for $G_\mu^\prime$ interactions with gluons [see 
Eq.~(\ref{self})] is given
in Appendix A of Ref.~\cite{Dobrescu:2007xf}. To leading order in $\alpha_s$, 
the partonic processes that lead to $G_\mu^\prime$ pair production at hadron colliders
have gluons (Fig.~\ref{fig:gg}) or 
quark-antiquark pairs (Fig.~\ref{fig:qqbar}) in the initial state.
The production cross section may be computed at tree level 
using {CalcHEP}~\cite{Pukhov:2004ca}. % with our model files \cite{web}.
For the gluon-gluon initial state, we find (in agreement with Ref.~\cite{Dicus:1994sw})
\bear
\sigma(gg \to\! G_\mu^\prime G_\mu^\prime) & \!\!
 = \!\! & \frac{9\pi \alpha_s^2}{16 \hat{s}^3} 
\left[ \beta \hat{s} \left( \frac{8 \hat{s}^2}{M_G^2} + 13 \hat{s} + 34 M_G^2 \right) \right.
\nonumber \\
&& \hspace{-5em} - \, \left.
8 \left(\hat{s}^2 + 3 M_G^2 \hat{s} - 3 M_G^4 \right) 
\ln \left( \frac{1 + \beta}{1 - \beta} \right) \right] ~, 
\label{sigma-gg}
\eear
where $\beta = (1 - 4 M_G^2/\hat{s})^{1/2}$ is the boost of $G_\mu^\prime$,
and $\hat{s}$ is the center-of-mass energy of the partonic collision.
This cross section is independent of $\hat{s}$ for 
$\hat{s} \gg M_G^2$ as a consequence of spin-1 exchange in the $t$ and $u$ channels.
Unitarity is preserved in this process independent of the gauge symmetry 
breaking sector because the radial modes of the $\Phi$ field or 
whatever else unitarizes longitudinal $G_\mu^\prime G_\mu^\prime$ scattering 
do not contribute to $gg \to G_\mu^\prime G_\mu^\prime$. Note that Eq.~(\ref{sigma-gg}) 
has the same large-$\hat{s}$ behaviour as the cross section for the standard model process 
$\gamma\gamma \to W^+W^-$ \cite{Belanger:1992qi}.

%%%%
%%
%%%%%%

%%%%
%%%%%%
Assuming  negligible couplings of $G'_\mu$ to standard model quarks ($h_q \ll 1$),
the cross section for $q\overline{q} \to G_\mu^\prime G_\mu^\prime$ 
depends only on $M_G$ and on the masses of the vectorlike quarks exchanged in the
$t$ and $u$ channels. We take these to be of the order of or larger than 
$M_G$ so that the vectorlike quarks do not affect the  $G_\mu^\prime$ decays.
For vectorlike quark masses equal to $M_G$, 
the process with quark-antiquark initial state has a cross section
\vspace*{-2.5mm}
\bear
\sigma(q\overline{q} \to G_\mu^\prime G_\mu^\prime) 
& = & \frac{\pi \alpha_s^2}{27 \hat{s}^2} 
\left[ \rule{0mm}{5mm} - \beta \left( 83 \hat{s} + 72 M_G^2 \right) \right.
\nonumber \\
&& \hspace{-5em} + \, \left.
2 \left(20 \hat{s} + 49 M_G^2 \right) 
\ln \left( \frac{1 + \beta}{1 - \beta} \right) 
\right] ~. 
\label{sigma-qqbar}
\eear 
This agrees with the cross sections 
obtained from the squared matrix elements computed in \cite{Macesanu:2002db}.
For vectorlike quark masses ($M_Q$) larger than $M_G$ or $\sqrt{\hat{s}}$ we find 
that $\sigma(q\overline{q} \to G_\mu^\prime G_\mu^\prime) \approx  \pi \alpha_s^2 \hat{s}/(18 M_G^4)$ 
up to corrections of order $1/M_Q^2$.
This cross section grows with $\hat{s}$ because the Goldstone boson eaten by 
$G_\mu^\prime$ has a coupling to $\bar{Q}\gamma^\mu T^a Q$ proportional to $M_Q$, so that $Q$ 
cannot be much heavier than $M_G$ if we keep $h_q \ll 1$.

%\smallskip
%%%%%%%%%%%%%%%%%%%%%%%%%%%%%%%%%%%%%%%%%%%%%%%%%%%%%%%%%%%%%%%%%%%%%%%%%
%{\it Signal and background at the Tevatron.}---

\section{\large \hspace*{-3mm} Signal and background at the Tevatron}

\begin{figure}[t]
\includegraphics[width=.47 \textwidth]{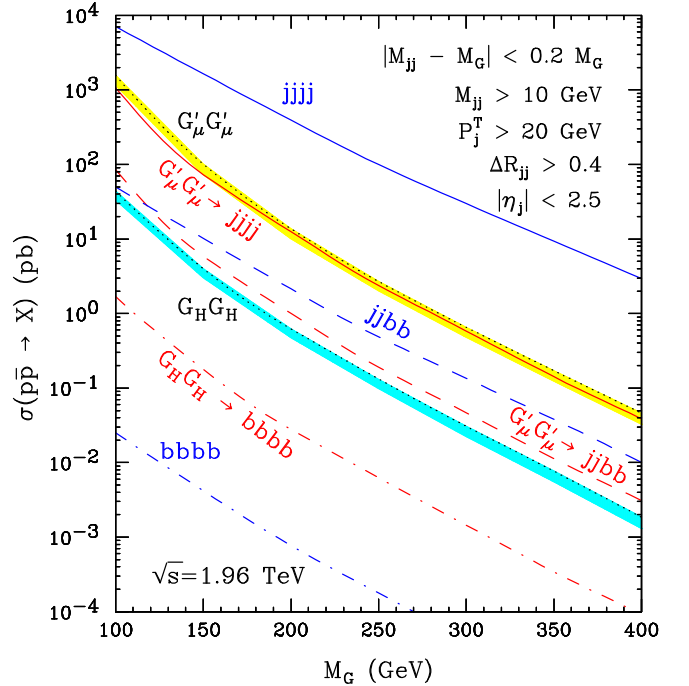}
\vspace*{-0.2cm}
\caption{\label{fig:sigmaGmu}
Tevatron cross sections for pair production of spin-1 ($G_\mu^\prime$) and spin-0 ($G_H$) octets 
and for backgrounds, as a function of octet mass. 
Four-jet final states  with 0, 2 and 4 $b$-tags are shown (for both signal and background)
as solid, dashed, and dot-dashed lines respectively, for given cuts.
Dotted lines represent signal cross sections without cuts, with uncertainties
indicated by shaded bands.}
\vspace*{-0.1cm}
\end{figure}

Taking the factorization and renormalization scale to be $\sqrt{\hat{s}}/2$, and 
using the CTEQ6L parton distributions~\cite{Pumplin:2002vw}, 
we obtain the leading-order cross section for $G_\mu^\one G_\mu^\one$ production
at the Tevatron shown in Fig.~\ref{fig:sigmaGmu}.
The shaded (yellow) band denotes the uncertainty in the cross section from varying 
the factorization and renormalization scale between $M_G$ and $\sqrt{\hat{s}}$. 
The uncertainty from varying $M_Q$ within the 
range $M_G$ to $2 M_G$ is even smaller. If vectorlike quarks are 
not included, then the cross section decreases (by a factor of two
in the perturbative window allowed by dijet searches, namely 
$h_q \approx 0.3$ and $M_G\approx 200$ GeV).
Next-to-leading order corrections are likely to be large
(they are of the order of 50\% for $t\bar{t}$ production \cite{Nason:1987xz},
and $G_\mu^\one$ has larger spin and color representation than the top quark),
but computing them  is beyond the scope of this letter. 

The main $G_\mu^\prime$ decays are into $q\bar{q}$, as the 
$G_\mu^\prime$ decays into gluons require higher-dimension operators 
which we neglect.
The signal due to the decay of a $G_\mu^\prime$ pair is 4 jets. 
Assuming equal branching fractions to all 
quark flavors, and given that decays to $t\bar{t}$ pairs are kinematically forbidden 
within the mass range accessible at the Tevatron, 
decays to 4 $b$-jets, $jjbb$, and 4$j$ will occur 4\%, 
32\% and 64\% of the time, respectively.

The dominant background is QCD multijet production.
Simulating this with both NJETS~\cite{Berends:1989ie} and 
MadGraph/MadEvent~\cite{Maltoni:2002qb} yielded consistent results of about 59 nb
for the cross section with standard cuts for jets (invariant mass $M_{jj} > 10$ GeV, 
transverse momentum $P_T > 20$ GeV, jet separation $\Delta R > 0.4$, and pseudorapidity $|\eta| < 2.5$).
This huge background can be dramatically reduced 
by requiring that the 4 jets form two dijet resonances with equal invariant mass \cite{Chivukula:1991zk}.
Of all possible pairings of the four jets, we 
choose the one with two pairs closest in invariant mass, thus reducing combinatorial backgrounds.
Since the intrinsic width of $G_\mu^\prime$ is small (its couplings to quarks
are assumed to be smaller than $g_s$), 
the width of the visible resonance is set by 
the detector resolution. Hence, an invariant mass cut of 
$|M_{jj} - M_G| < 0.2 M_G$ suppresses the background by orders of magnitude  
(see the curve labelled $jjjj$ in Fig.~\ref{fig:sigmaGmu}).
Final states involving $b$-jets have even smaller background, but their rates
are also suppressed by the $b$-tagging efficiency, which we assume to be 50\%  
(the $b$ mistag rate is taken to be 1\%).  
Fig.~\ref{fig:sigmaGmu} indicates
that CDF and {D\O} could make a 5$\sigma$ discovery of a $G^\prime_\mu$ of
mass below approximately 340 GeV (320 GeV) in the  $jjbb$ (4$j$) final state 
with an integrated luminosity of $4$ fb$^{-1}$.  The significance increases
fast for lower masses ($40\sigma$ at $M_G=200$ GeV), allowing flexibility in the 
choice of cuts. Note that although the QCD background does not include
resonant structures, the cuts could induce a false signal 
with two dijet resonances \cite{Chivukula:1991zk}. 
However, this false signal can be eliminated
because changes in cuts would shift the position of the cut-induced resonances without 
affecting the location of the resonances due to $G^\prime_\mu$ decays. 
It is likely though that a detailed analysis that includes systematic 
errors on background would yield a smaller significance, especially at low invariant mass.

%%%%%%%%%%%%%%%%%
%

%\smallskip 
%%%%%%%%%%%%%%%%%%%%%%%%%%%%%%%%%%%%%%%%%%%%%%%%%%%%%%%%%%%%%%%%%%%
\section{ \large   Spin-0 color octet}

We now discuss the case of spin-0 octets.
As a consequence of $SU(3)_c$ gauge invariance, at tree level $G_H$ 
couples to gluons only in pairs (see Eq. (2.4) of \cite{Dobrescu:2007xf}).
In general, there are no renormalizable interactions of the spinless gluon 
to standard model fermions (only if the color octet is an
$SU(2)_W$ doublet and carries hypercharge $\pm 1/2$ does it have Yukawa interactions with 
quarks \cite{Manohar:2006ga}).
However, the following dimension-5 operator can exist
\vspace*{-2mm}
\be
\label{equ:GHqq}
\frac{i c_s}{M_{G}}\left(\overline{q}\gamma^\mu \gamma_5 T^a q \right)  \partial_\mu G_H^a ~,
\vspace*{-2mm}
\ee 
where $c_s$ is a dimensionless parameter.  For simplicity we ignore flavor
off-diagonal operators.
The operator (\ref{equ:GHqq}) can be induced in a renormalizable theory that includes 
tree-level exchange of either a vectorlike quark 
or a weak-doublet color-octet scalar,
or it may be induced at loop level provided that $G_H$ has additional interactions, 
such as a cubic self-coupling. 
After integration by parts and use of the field equation for the quark,
the operator  (\ref{equ:GHqq}) becomes proportional to the quark mass.
As a result, $G_H$ decays into $q\bar{q}$ pairs with widths proportional to the quark mass squared. 
If the coefficient $c_s$ is flavor independent,
$G_H$ decays predominantly into $b\bar{b}$, or $t\bar{t}$ 
for masses above 350 GeV.
However, it is possible that $c_s$ is nonzero only
for down-type quarks, so that even above the  $t\bar{t}$ threshold the dominant decay is into
$b\bar{b}$ (conversely, $G_H$ could couple exclusively to up-type quarks).

$G_H$ may be produced only in pairs, and the 
tree-level partonic processes at hadron colliders
are analogous to the ones shown in Figs.~\ref{fig:gg} and~\ref{fig:qqbar},
with $G^\one_\mu$ replaced by $G_H$ (vectorlike quarks are not included here).
The partonic cross sections are given in  \cite{Chivukula:1991zk,Manohar:2006ga}. 
In Fig.~\ref{fig:sigmaGmu} we 
show the leading-order $G_H G_H$ production cross section at the Tevatron, and the 
$4b$ signal after cuts, taking the branching fraction of $G_H \to b\bar{b}$ to be 100\%.
A 5$\sigma$ discovery of a $G_H$ with at least 10 signal events can be made 
for mass below 280 GeV in the $4b$ final state with $4$ fb$^{-1}$.
The spin of the octet can be determined  using angular distributions, 
as the events can be fully reconstructed in the center-of-mass frame. 
\begin{figure}[tbp]
\includegraphics[width=.47 \textwidth]{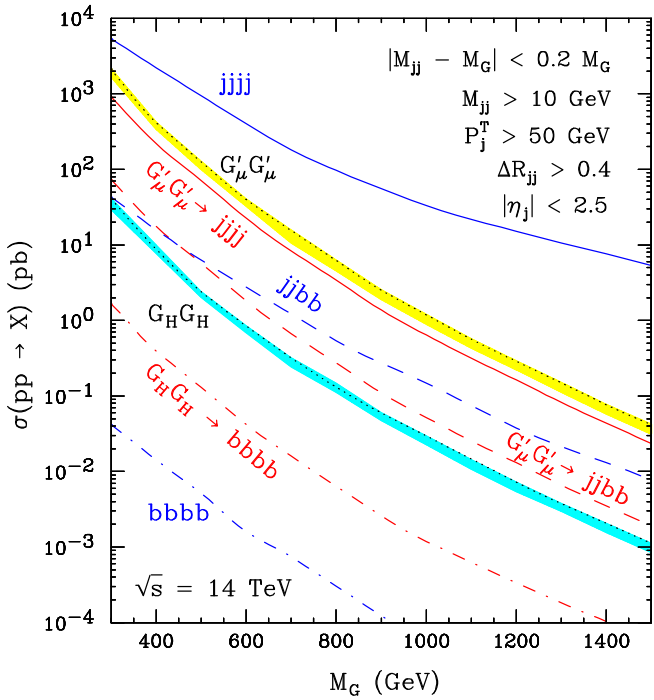}
\vspace*{-0.2cm}
\caption{\label{fig:sigmaLHC}
Same as Fig.~\ref{fig:sigmaGmu}, for the LHC. }
%The branching fraction of $G_H \to b\bar{b}$ is taken to be 100\%.}
\vspace*{-0.1cm}
\end{figure}
%\smallskip
%%%%%%%%%%%%%%%%%%%%%%%%%%%%%%%%%%%%%%%%%%%%%%%%%%%%%%%%%%%%%%%%%%%%%%%%%

%\hspace*{-4mm}
%{\it Color octets at the LHC.}---

\section{ \large   Color octets at the LHC}

Production cross sections for octets at the LHC are dominated 
by the gluon initial state~\cite{Dobrescu:2007xf,Manohar:2006ga}, 
making them almost entirely model-independent. 
We plot these in Fig.~\ref{fig:sigmaLHC}, for equal branching fractions 
of $G_\mu^\prime$ to all quark flavors 
and a 100\% branching fraction of $G_H$ to $b\bar{b}$ %(NLO
(next-to-leading order 
corrections, which may be sizable, are not included).
The production cross section is huge, more than 20 times that for a quark of equal mass,
allowing for early discovery.
Repeating our Tevatron analysis with a stronger $P_T$ cut of 50 GeV at the LHC, we find
a 5$\sigma$ reach in the $4 j$ and $jjbb$ final states for $G_\mu^\prime$ of mass
below 0.92 and 0.96 TeV (1.2 and 1.3 TeV), 
respectively, with 1 fb$^{-1}$ (10 fb$^{-1}$);
in the $4b$ channel, requiring 5$\sigma$ and at least 10 events,
the $G_H$ mass reach is 0.75 TeV (1.0 TeV).

%%%%%%%%%%%%%%%%%%%%%%%%%%%%%%%%%%%%%%%%%%%%%%%%%%%%%%%%%%%%%%%%%%%%
%
\begin{figure}[t]
\includegraphics[width=.465 \textwidth]{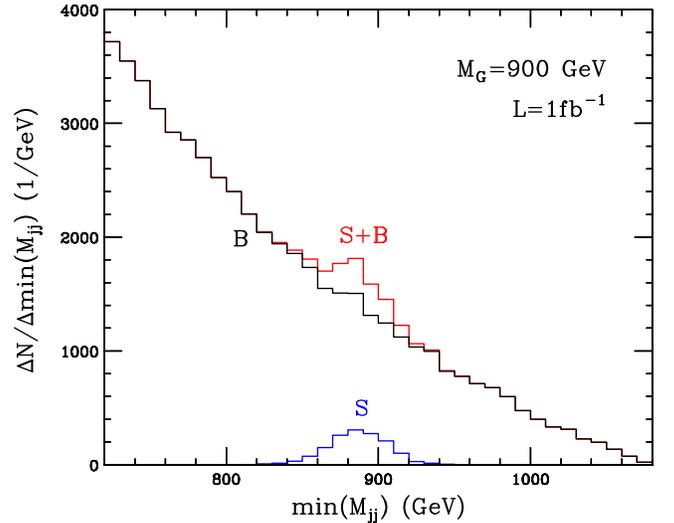}
\vspace*{-0.5cm}
\caption{\label{fig:invmass}
Invariant mass distributions for background (B) and signal (S) at the LHC, for $M_G = 900$ GeV,
as a function of the smaller of the two dijet masses.  
The dijets chosen have the closest invariant masses of all
possible pairings of the four jets in the final state.
%The two dijets chosen are the ones closest in invariant mass out of all 
%possible pairings of the four jets in the final state.
%for background (B) and signal (S) for $M_G = 900$ GeV. 
% The four jets are paired
% such that the two dijets are closest in invariant mass.
%Invariant mass distributions as a function of the smaller of the two dijet masses, 
%for background (B) and signal (S) for $M_G = 900$ GeV.
The number of events is normalized to a luminosity of 1 fb$^{-1}$.}
\vspace*{-0.1cm}
\end{figure}
%\smallskip 
%%%%%%%%%%%%%%%%%%%%%%%%%%%%%%%%%%%%%%%%%%%%%%%%%%%%%%%%%%%%%%%%%%%

Given the large background, one might be concerned that the signal is hard to  isolate.
We have checked that generally the background distribution falls rapidly with the invariant dijet mass.
In some cases a peak may appear near the boundary of the invariant mass distribution, but
it is easily eliminated by changing the cuts or binning.
In Fig.~\ref{fig:invmass} we show the invariant mass distribution, after cuts,
for background and signal in the $4j$ channel for $M_G= 900$ GeV and  a 
luminosity of 1 fb$^{-1}$. 
%We generated parton-level events for background and signal using MadGraph/MadEvent~\cite{Maltoni:2002qb} 
%and {CalcHEP}~\cite{Pukhov:2004ca}, respectively.
We simulated the response of the hadronic calorimeter using Gaussian smearing with an energy 
resolution given by \cite{:1999fr}:
%The energy resolution of the hadronic calorimeter was taken to be \cite{:1999fr}:
%
\begin{equation}
\left(\frac{\Delta E}{E}\right)^{\! 2} = \frac{0.5^2}{E ~({\rm GeV})} + 0.03^2 \, .
\end{equation}
%
%We then applied cuts defined in Fig.~\ref{fig:sigmaLHC} 
%to select a pair of dijet out of 3 possible combinations.
This 6$\sigma$  excess is clearly visible over the background for the smaller of the two dijet masses, ${\rm min}(M_{jj})$. The shift of the peak to a slightly lower mass is due to smearing, and may be corrected, once an excess is observed, by using an alternative variable such as the average of the two dijet masses.

%%%%%%%%%%%%%%%%%%%%%%%%%%%%%%%%%%%%%

For $G_\mu^\prime$ or $G_H$ mass sufficiently above the $t\bar{t}$ threshold
there are additional signals: two $t\bar{t}$ resonances \cite{Dicus:1994sw},
or a $t\bar{t}jj$ final state with $t\bar{t}$ and $jj$ separately 
reconstructed as resonances of same mass. Note also that if the octet has flavor off-diagonal
couplings, then other pairs of resonances would be seen ($t\bar{t}$ +  $tj$, $tj$ +  $tj$, or
$jj$ +  $tj$, and similar combinations involving $b$ jets).

To conclude, pair production of heavy color-octet bosons at the Tevatron and LHC 
is copious and mostly model-independent. It leads to spectacular signatures with jets and $b$-jets, 
and possibly top quarks, that reconstruct two narrow resonances of equal mass.  Hence the substantial QCD 
backgrounds are greatly reducible, and dedicated searches can discover such particles for a large range of masses.

%\vspace*{-7mm}

%%%%%%%%%%%%%%%%%%%%%%%%%%%%%%%%%%%%%%%%%%%%%%%%%%%%%%%%%%%%%%%%%%%
%%%%%%%%%%%%%%%%%%%%%%%%%%%%%%%%%%%%%%%%%%%%%%%%%%%%%%%%%%%%%%%%%%%
 \vfil 
\end{document}